# Testing for universality of Mendeley readership distributions

Ciriaco Andrea D'Angelo
dangelo@dii.uniroma2.it

and

Samuele Di Russo
samuele995@hotmail.it

*University of Rome "Tor Vergata"*
*Dept of Engineering and Management*
*Via del Politecnico 1, 00133 Rome - ITALY*

**Abstract**

Altmetrics promise useful support for assessing the impact of scientific works, including beyond the scholarly community and with very limited citation windows. Unfortunately, altmetrics scores are currently available only for recent articles and cannot be used as covariates in predicting long term impact of publications. However, the study of their statistical properties is a subject of evident interest to scientometricians. Applying the same approaches used in the literature to assess the universality of citation distributions, the intention here is to test whether the universal distribution also holds for Mendeley readerships. Results of the analysis carried out on a sample of publications randomly extracted from the Web of Science confirm that readerships seem to share similar shapes across fields and can be rescaled to a common and universal form. Such rescaling results as not particularly effective on the right tails. In other regions, rescaling causes a good collapse of field specific distributions, even for very recent publications.

# 1. Introduction

The essence of scientific activity is information processing: the science system consumes, transforms, produces, and exchanges "information". Scientists talk to one another, read each other's papers, and most importantly, they publish scientific papers. Bibliometrics is the study of measuring and analyzing the information encoded in published papers. Scientists collect and analyze prior knowledge, add value to it producing new knowledge, which they nearly always encode in papers made accessible to other scientists, and so contribute to further scientific and technical advancement. For the production of new knowledge to have an impact "on scientific advancement", it has to be used by other scientists (Abramo, 2018). Scientometricians generally consider citations as "certification" of the use of prior knowledge: such assumption is based on the belief that scientists normally cite papers to recognize their influence (Kaplan, 1965; Merton, 1973), although there could also be exceptions (uncitedness, negative citations, undercitation, overcitation, etc.).

Citation-based analysis, however, is unable to capture impact outside the scientific system, such as on practitioners or educators. Moreover, citation count can be considered as a reliable proxy of the scholarly impact of a work only when there is an adequate "citation time window": the problem then being that to serve its purpose, no performance assessment aimed at truly informing policy and management decisions could await the decades for completion of the citation life-cycle.

These two considerations – the tradeoff between timeliness and robustness of impact evaluation, and the ability to observe impact even outside the scholarly community – bring about the evident interest of scientometricians in the so-called "altmetrics" (Fenner, 2014; Sugimoto, Work, Larivière, & Haustein, 2017; Thelwall, 2017a; Thelwall, 2017b). Known mainly for their ease of use and the wide range of indicators, many scholars are persuaded that altmetrics offer an alternative view of impact, compared to traditional bibliometric indicators.

Previous literature found positive but relatively weak correlations between altmetrics and citations, thus supporting the idea that altmetrics do not reflect the same concept of impact as citations, i.e. the real "use" of the cited/read publication (Costas, Zahedi, & Wouters, 2015).

Nevertheless, altmetrics promise useful support for assessing the impact of scientific work: reading a publication occurs before citing it, so it could make sense to count "readers" rather than (or in addition to) citations for recent publications (Priem, Taraborelli, Groth, & Neylon, 2010; Li, Thelwall, & Giustini, 2012; Thelwall, Haustein, Larivière, & Sugimoto, 2013; Shema, Bar-Ilan, & Thelwall, 2014; Sud, & Thelwall, 2014; Thelwall & Sud, 2016). In other words, altmetrics could serve as covariates, to improve predictive power of early citations when the time window is too short (Shema, Bar-Ilan, & Thelwall, 2014; Abramo, D'Angelo, Felici, 2018). Unfortunately, the altmetrics serving such purposes are still widely available only for recent articles. Thelwall (2018) compares reader counts within a month of publication for ten fields, with citation counts after 20 months: a window too short to observe "real" impact. Leaving aside the so-called "sleeping beauties", observation of real impact generally requires citation windows of a few decades[1]: consider for example a window of 20 years, i.e.

---
[1] Here we are referring to the very lifecycle of a publication, not to a citation window suitable for predicting its total citations (Stringer, Sales-Pardo & Nunes Amaral, 2008; Wang, 2013).

works published at the end of the last century, for which no early altmetric data are available (altmetric counts at 2000, 2001 or 2002), but only aggregate data to date.

However, it would be interesting to study the statistical properties of the altmetrics for purposes of optimizing their use in quantitative analyses, meaning the way citation-based indicators are already applied for assessing publications, researchers, research teams, departments, universities, or even countries. Given the varying intensity of publications (and of citations) between fields, the citation-based indicators used in such applications must be standardized. The use of altmetrics as an alternative or as a covariate of citations then implies the same process, which in turns calls for deeper understanding of the statistical properties of the reference distributions.

The current study aims at verifying the existence of similarities between citation patterns and patterns of Mendeley readership counts, with particular emphasis on the "universality" of rescaled distributions, as claimed in the literature for citations but still unexplored for altmetrics. Specifically, we aim to answer the following research questions:

- Do readership distributions share the same shape across fields?
- Can such distributions be rescaled to a common and universal form?
- Is there empirical evidence of such universality in all regions of readership distributions, and in particular in their right tails?

To do this, we: i) extract a sample of 2009-2016 publications indexed in the Web of Science Core Collection; ii) cluster them based on publication year and subject category (SC) of the hosting journal, and; iii) analyze the distributions of Mendeley readerships for each cluster.

In order to answer the above research questions we applied: i) a qualitative graphic comparison of original and rescaled distributions; ii) the Characteristic Scores and Scales method proposed by Glänzel and Schubert (1988).

The next section provides a summary of previous literature on the topic. Section 3 illustrates the empirical dataset and methodology; Section 4 shows the results of the analyses and Section 5 concludes the work, summarizing the main research findings and providing final remarks.

**2. Literature review**

Among the altmetrics currently available within systems for diffusion of scientific literature, the one receiving the most attention is certainly the Mendeley readership count (Priem, Piwowar, & Hemminger, 2012; Zahedi, Costas & Wouters, 2014). The platform of the same name records the activities of "reading" of scientific literature by profiled users. This indicator shows good correlation with citation count (Costas, Zahedi, & Wouters, 2015), although there is a remarkable difference between the rate of accumulation of readerships and that for citations (Costas, Zehedi, & Wouters, 2015; Thelwall, & Sud, 2016).

In turn, citation distributions are typical of each field, and in all case are particularly skewed. Citation distributions have very different characteristics across sub-fields, a mean significantly higher than the median, and a limited share of articles accounting for a large share of citations (Albarrán, Crespo, Ortuño, & Ruiz-Castillo, 2011; Bornmann & Leydesdorff, 2017).

Ruiz-Castillo (2012) confirms that citation distributions share the same broad shape and are highly skewed, adding that they are often crowned by a power law. For Franceschet (2011), in computer science, the skewness registered for conference publications is also more pronounced than that for journal papers.

Thelwall (2016) found that the hooked power law fits well for natural, life and medical sciences while the discretised lognormal distribution proves to be better for the arts&humanities, social sciences, engineering and technology areas. Stringer, Sales-Pardo & Amaral (2008) confirm that citations to articles in a given year and subject category are likely to be approximately log-normally distributed.

A theme of interest to many scholars is that the probability distributions of citations can be rescaled to a common form, both at the level of individual publications (Radicchi, Fortunato, & Castellano, 2008) and of journals or institutions (Chatterjee, Ghosh, & Chakrabarti, 2016).

Radicchi, Fortunato, and Castellano (2008), referring to articles published in 1999 and 2004 in journals of 14 different Web of Science (WoS) subject categories (SCs), demonstrate that the citation distributions can be rescaled to a "universal" form, considering the average number of citations of all articles in the same SC and publication year. Confirmation of the universality claim also arrives from other studies by the same authors, conducted on fields of Physics (Radicchi & Castellano, 2011) and other sciences (Radicchi & Castellano, 2012), as well as from Bornmann and Daniel (2009) for Chemistry.

Waltman, van Eck and Van Raan (2012), in an analysis extending to all 221 categories of sciences and social sciences, provide only partial confirmation of the universality of the citation distributions: in the SCs with lower citation intensity, the distribution does not always follow the log-normal model. The authors note, however, that "the universality claim becomes more justifiable when uncited publications are excluded from the analysis." Albarrán, Crespo, Ortuño, and Ruiz-Castillo (2011) also report that when looking "into subsets of articles within the lower and upper tails of citation distributions the universality partially breaks down". As for the upper tail, "outliers" are by definition difficult to treat and to associate with a universal shape. For the lower tail, uncited publications are critical and their treatment is "not easy": their share varies widely between fields and different scalings imply different transformation of original distributions. However, these observations do not automatically imply that the universality claim is always "unjustifiable". In fact, Waltman et al. (2012) observe that the use of rescaled bibliometric indicators is still justified when citation distributions are not universal, a position shared by Ruiz-Castillo (2012). In this regard, Abramo, Cicero, and D'Angelo (2012) confirm that for optimum rescaling, uncited publications should be excluded from calculations.

The method known as Characteristic Scores and Scales, introduced by Glänzel and Schubert (1988) investigates citation patterns from a different perspective: regardless of the field and citation window, class distribution follows a stable and proportionate pattern: 69-70% of works belong to the "poorly" cited class; 21% to the "fairly" cited class; 6-7% are "remarkably" cited; 2-3% are "outstandingly" cited. Albarrán, Crespo, Ortuño, and Ruiz-Castillo (2011) revealed that when analyzed with the CSS technique, the shape of citation distributions for publications in different fields appears strikingly similar. Vîiu (2018) argues that the pattern defined by the CSS method is directly attributable to the universality of citation distributions, by virtue of their log-normal character, most clearly observed in the fields with the highest citation rate. The objective of the current work is to investigate the statistical properties of readership counts, in search of the same "character" of universality for Mendeley readership distributions. At the moment the only study addressing such an issue is that by Thelwall and Wilson (2016). Based on a Scopus dataset related to 45 medical fields they confirm that the Mendeley readerships follow lognormal distributions in many fields, while the hooked power law fits citation data appropriately in some other fields.

The explosion of interest in altmetrics calls for the extension of such analysis to non-medical fields, and examination of whether the readerships can be rescaled to a common and universal form. Such analysis would be especially timely, given the attraction of using altmetrics in substitution or as a supplement to citations, for measuring the impact of scientific publications of different years and in different scientific domains.

## 3. Data and sources

The dataset for the analysis is a sample of publications indexed in WoS in the period 2009-2016. For each year, a total of 10,000 publications (only "article" document type) were randomly extracted from the WoS Core collection, selecting only from those with publication dates in December. The latter condition ensures precise definition of the citation window of the articles, a particularly important requirement in the case of more recent ones. Data matching from WoS to Mendeley turned out to be quite laborious and this forced us to adopt a sample whose size is a trade-off between the need for completeness on one hand, and budget constraints on the other.

The total 80,0000 extracted publications were then classified in the relevant field through the WoS SCs associated with the journal. From these, the selection was further limited to the top 30 SCs by number of WoS-indexed publications over the entire period, leaving a total of 42,291 publications. This choice ensures adequate sample size for each SC-year combination and good representation of the macro scientific domain, without introducing systematic bias.

The publications were matched to the corresponding records in Mendeley using Webometric Analyst 2.0 software, which incorporates the Mendeley open API.[2] To maximize accuracy, only the matches between WoS articles and Mendeley records with probability greater than 90% were considered correct. Of the 42,291 sampled publications, 98.2% (40,632 items) were matched to a corresponding Mendeley record: the breakdown by SC is shown in Table 1.[3] With the exception of Economics, social science fields are omitted from the dataset, as well as all arts and humanities subject categories.

For each article in the dataset, citations in WoS and readerships in Mendeley were retrieved at 1 February 2018.

---

[2] A tool developed by the Statistical Cybermetrics Research Group, led by Professor Mike Thelwall at University of Wolverhampton. For details, http://lexiurl.wlv.ac.uk/, last accessed 20 March 2019. The software and matching algorithm operations are illustrated in Thelwall and Wilson (2016).
[3] WoS articles not matched in Mendeley were excluded from the dataset.

*Table 1: Dataset of the analysis, by SC*

| Subject category | Total 2009-2016 WoS records | Sampled articles | Matched in Mendeley |
|---|---|---|---|
| Materials science, multidisciplinary | 583,037 | 2,860 (0.5%) | 2,804 (98.0%) |
| Biochemistry & molecular biology | 447,901 | 3,076 (0.7%) | 3,021 (98.2%) |
| Chemistry, physical | 442,255 | 2,892 (0.7%) | 2,868 (99.2%) |
| Chemistry, multidisciplinary | 427,408 | 1,951 (0.5%) | 1,880 (96.4%) |
| Physics, applied | 415,330 | 2,258 (0.5%) | 2,230 (98.8%) |
| Engineering, electrical & electronic | 401,914 | 1,629 (0.4%) | 1,572 (96.5%) |
| Environmental sciences | 307,178 | 1,713 (0.6%) | 1,643 (95.9%) |
| Neurosciences | 290,357 | 1,909 (0.7%) | 1,831 (95.9%) |
| Oncology | 283,913 | 1,584 (0.6%) | 1,458 (92.0%) |
| Surgery | 264,321 | 1,162 (0.4%) | 1,044 (89.8%) |
| Pharmacology & pharmacy | 255,544 | 1,252 (0.5%) | 1,162 (92.8%) |
| Nanoscience & nanotechnology | 223,720 | 1,330 (0.6%) | 1,319 (99.2%) |
| Physics, condensed matter | 218,798 | 1,171 (0.5%) | 1,161 (99.1%) |
| Public, environmental & occupational health | 217,162 | 1,103 (0.5%) | 1,024 (92.8%) |
| Engineering, chemical | 216,299 | 925 (0.4%) | 877 (94.8%) |
| Mathematics | 210,960 | 854 (0.4%) | 767 (89.8%) |
| Biotechnology & applied microbiology | 207,186 | 1,032 (0.5%) | 991 (96.0%) |
| Optics | 205,724 | 1,172 (0.6%) | 1,110 (94.7%) |
| Mathematics, applied | 204,299 | 652 (0.3%) | 600 (92.0%) |
| Cell biology | 202,239 | 1,444 (0.7%) | 1,400 (97.0%) |
| Clinical neurology | 194,394 | 1,008 (0.5%) | 959 (95.1%) |
| Energy & fuels | 182,706 | 947 (0.5%) | 921 (97.3%) |
| Economics | 181,619 | 1,590 (0.9%) | 1,517 (95.4%) |
| Physics, multidisciplinary | 180,604 | 1,164 (0.6%) | 1,106 (95.0%) |
| Plant sciences | 172,359 | 740 (0.4%) | 650 (87.8%) |
| Chemistry, analytical | 171,417 | 794 (0.5%) | 772 (97.2%) |
| Food science & technology | 166,293 | 771 (0.5%) | 717 (93.0%) |
| Immunology | 166,228 | 1,116 (0.7%) | 1,071 (96.0%) |
| Genetics & heredity | 161,543 | 1,061 (0.7%) | 1,046 (98.6%) |
| Microbiology | 160,390 | 1,131 (0.7%) | 1,111 (98.2%) |
| Total | 7,763,098 | 42,291 (0.5%) | 40,632 (96.1%) |

## 4. Results and analysis

### 4.1 Original and rescaled readership distributions

As seen in Section 2, previous literature indicates that citation distributions share the same shape, being highly skewed and often crowned by a log-normal power law, or hooked power law model. Analysing publications from 45 medical fields, Thelwall (2016) found that the readership data hold a highly skewed distribution as well, generally fitted to a lognormal, but in some fields a hooked power law has proven to be more appropriate. Our intention is to now extend this analysis to our dataset in order to verify if readerships share similar patterns, and whether the rescaled distributions, obtained with the same method presented by Radicchi, Fortunato, and Castellano (2008) for citations, have a "universal character".

Table 2 presents some statistics for 2010 publications in each SC.[4] Column 3 reveals a high variability in average number of readerships per publication ($R_0$) across SCs, ranging from 6.2 in Mathematics to 44.7 in Cell biology. Extreme values ($R_{max}$) also show evident field dependence: the most read article in Biotechnology & applied microbiology registers a number of reads (1649) ten times that for the corresponding article in Mathematics (17).

---
[4] See Appendix A for complete data of all publication years.

The fifth column shows the Shapiro-Wilk (SW) test p-value for log-normality, one of the most powerful tests for verifying normality, especially for small samples and non-aggregated data (Shapiro & Wilk, 1965).[5] Its utilisation finds justification given the varying size of samples in the dataset, some of which are quite small (59 observations for "plant sciences" 2009). The SW test reveals that the null hypothesis of data log-normality cannot be rejected in 29 of the 30 investigated SCs (the exception is Mathematics). Finally, each distribution was fitted with a lognormal model:[6]

$$F(r) = \frac{1}{\sigma r \sqrt{2\pi}} e^{-\frac{[\log(r)-\mu]^2}{2\sigma^2}}$$

[1]

The right side of table 3 shows the fitting values of such model ($\eta$ and $\sigma^2$):
- all fitted values of $\eta$ reported in the table are compatible within 2 standard deviations, except for Economics and Engineering, electrical & electronic (compatible within 3 standard deviations);
- all fitted values of $\sigma^2$ are compatible within 1 standard deviation.

According to previous literature, the variability in distribution of citations observed at the field level can be related to the pattern of citations in the field, measured by the average number of citations per article (Moed, De Bruin, & Vanleeuwen, 1995; Rehn, Kronman, & Wadsko, 2007; Radicchi, Fortunato, & Castellano, 2008; Abramo, Cicero, & D'Angelo, 2012). This justifies, *inter alia*, the adoption of relative indicators where the impact of a publication is measured by the ratio between the number of citations received and the average number of citations received by articles published in its field in the same year.

We can assume that the same holds true for readerships, i.e. that differences in readership distributions are essentially due to specific patterns in the subject category, as measured by the average number of readership per publications in the same year and subject category ($R_0$).

We can expect that in rescaling the distribution of readerships for publications in the same SC by their average, a universal curve could be found.

---

[5] With Bonferroni correction considering m=240 hypotheses and a desired alpha = 0.05.
[6] Here we are not interested in comparing multiple models, or testing prior hypothesis that one model is "correct"; based on main results obtained by Thelwall (2016), we used log-normal model only for contrasting distributions related to different fields.

*Table 2: Statistics of readerships for 2010 publications, by subject categories and ML fit of lognormal distribution*

| Subject category | Obs | $R_0$ | $R_{max}$ | SW_Prob>z | $\mu$ Coeff. | z | P>|z| | $\sigma^2$ Coeff. | z | P>|z| | LogLik |
|---|---|---|---|---|---|---|---|---|---|---|---|
| Biochemistry & molecular biology | 412 | 29.6 | 739 | 0.883 | -0.594 | -11.26 | 0.000 | 1.11 | 28.71 | 0.000 | -365.5 |
| Biotechnology & applied microbiology | 166 | 27.6 | 1649 | 0.913 | -0.589 | -7.35 | 0.000 | 1.10 | 18.22 | 0.000 | -139.9 |
| Cell biology | 181 | 44.7 | 739 | 0.909 | -0.761 | -8.24 | 0.000 | 1.12 | 19.03 | 0.000 | -162.3 |
| Chemistry, analytical | 106 | 24.5 | 183 | 0.915 | -0.525 | -4.83 | 0.000 | 1.11 | 14.56 | 0.000 | -100.7 |
| Chemistry, multidisciplinary | 195 | 35.9 | 374 | 0.900 | -0.720 | -8.16 | 0.000 | 1.12 | 19.75 | 0.000 | -192.0 |
| Chemistry, physical | 399 | 25.3 | 301 | 0.874 | -0.542 | -10.01 | 0.000 | 1.11 | 28.25 | 0.000 | -382.4 |
| Clinical neurology | 99 | 24.7 | 201 | 0.918 | -0.720 | -5.94 | 0.000 | 1.12 | 14.07 | 0.000 | -93.2 |
| Economics | 175 | 27.5 | 1337 | 0.910 | -1.068 | -10.15 | 0.000 | 1.14 | 18.71 | 0.000 | -122.0 |
| Energy & fuels | 113 | 27.2 | 242 | 0.918 | -0.502 | -4.86 | 0.000 | 1.11 | 15.03 | 0.000 | -105.6 |
| Engineering, chemical | 119 | 19.9 | 350 | 0.912 | -0.638 | -5.97 | 0.000 | 1.12 | 15.43 | 0.000 | -112.0 |
| Engineering, electrical & electronic | 103 | 16.3 | 135 | 0.899 | -0.968 | -7.26 | 0.000 | 1.14 | 14.35 | 0.000 | -91.2 |
| Environmental sciences | 200 | 25.4 | 486 | 0.904 | -0.469 | -6.65 | 0.000 | 1.00 | 20.00 | 0.000 | -186.6 |
| Food science & technology | 108 | 16.5 | 141 | 0.910 | -0.432 | -4.48 | 0.000 | 1.10 | 14.70 | 0.000 | -110.7 |
| Genetics & heredity | 159 | 31.6 | 1648 | 0.913 | -0.877 | -8.87 | 0.000 | 1.13 | 17.83 | 0.000 | -128.7 |
| Immunology | 142 | 28.8 | 141 | 0.900 | -0.504 | -5.54 | 0.000 | 1.11 | 16.85 | 0.000 | -139.4 |
| Materials science, multidisciplinary | 343 | 27.3 | 301 | 0.878 | -0.571 | -9.84 | 0.000 | 1.11 | 26.19 | 0.000 | -326.8 |
| Mathematics | 85 | 6.2 | 17 | 0.001 | -0.396 | -4.18 | 0.000 | 0.88 | 13.04 | 0.000 | -76.4 |
| Mathematics, applied | 75 | 8.7 | 60 | 0.603 | -0.540 | -4.53 | 0.011 | 1.10 | 12.25 | 0.000 | -68.3 |
| Microbiology | 111 | 23.7 | 205 | 0.911 | -0.415 | -4.62 | 0.000 | 0.95 | 14.90 | 0.000 | -111.3 |
| Nanoscience & nanotechnology | 183 | 32.8 | 178 | 0.900 | -0.507 | -6.46 | 0.000 | 1.11 | 19.13 | 0.000 | -176.4 |
| Neurosciences | 190 | 28.2 | 306 | 0.904 | -0.474 | -6.37 | 0.000 | 1.10 | 19.49 | 0.000 | -181.1 |
| Oncology | 208 | 34.5 | 348 | 0.894 | -0.650 | -7.90 | 0.000 | 1.12 | 20.40 | 0.000 | -197.6 |
| Optics | 143 | 19.6 | 254 | 0.896 | -0.628 | -6.54 | 0.000 | 1.12 | 16.91 | 0.000 | -138.3 |
| Pharmacology & pharmacy | 134 | 19.7 | 213 | 0.893 | -0.503 | -5.51 | 0.000 | 1.11 | 16.37 | 0.000 | -118.4 |
| Physics, applied | 255 | 19.2 | 181 | 0.875 | -0.737 | -9.33 | 0.000 | 1.13 | 22.58 | 0.000 | -248.8 |
| Physics, condensed matter | 128 | 23.7 | 268 | 0.907 | -0.742 | -6.99 | 0.000 | 1.12 | 16.00 | 0.000 | -122.5 |
| Physics, multidisciplinary | 148 | 30.8 | 201 | 0.903 | -0.789 | -6.95 | 0.000 | 1.14 | 17.20 | 0.000 | -147.5 |
| Plant sciences | 64 | 21.3 | 308 | 0.936 | -0.700 | -4.54 | 0.010 | 1.12 | 11.31 | 0.000 | -54.4 |
| Public, environmental & occupational health | 93 | 20.0 | 235 | 0.926 | -0.679 | -5.71 | 0.000 | 1.12 | 13.64 | 0.000 | -88.1 |
| Surgery | 96 | 21.6 | 101 | 0.883 | -0.637 | -5.28 | 0.000 | 1.12 | 13.86 | 0.000 | -80.8 |

$R_0$= mean number of readerships; $R_{max}$= max number of readerships; SW = Shapiro Wilk log-normality test; $\mu$, $\sigma^2$ = fitting parameter of the lognormal curve

Hence, given the rescaled values of readerships ($r_r$) as:[7]

$$r_r = \frac{r}{R_0}$$

[2]

We proceed in collapsing all SC distributions (of $r_r$), obtaining a new general distribution. Table 3 reports the fitting parameter of equation [1] applied to this general distribution, for each publication year.

*Table 3: ML fit of lognormal distributions for rescaled readerships, by publication year*

| | | μ | | | σ² | | | |
|---|---|---|---|---|---|---|---|---|
| Year | Obs | Coeff. | Std. Err. | Prob > z | Coeff. | Std. Err. | Prob > z | LogLik |
| 2009 | 4874 | -0.538 | 0.015 | 0.000 | 1.056 | 0.011 | 0.000 | -4558.2 |
| 2010 | 4933 | -0.531 | 0.015 | 0.000 | 1.061 | 0.010 | 0.000 | -4672.3 |
| 2011 | 4668 | -0.533 | 0.015 | 0.000 | 1.050 | 0.011 | 0.000 | -4364.2 |
| 2012 | 4747 | -0.534 | 0.015 | 0.000 | 1.059 | 0.011 | 0.000 | -4476.5 |
| 2013 | 4663 | -0.443 | 0.014 | 0.000 | 0.975 | 0.010 | 0.000 | -4433.7 |
| 2014 | 5072 | -0.487 | 0.014 | 0.000 | 1.010 | 0.100 | 0.000 | -4779.1 |
| 2015 | 4562 | -0.503 | 0.015 | 0.000 | 1.010 | 0.010 | 0.000 | -4225.5 |
| 2016 | 3250 | -0.412 | 0.017 | 0.000 | 0.940 | 0.012 | 0.000 | -3070.8 |

For all eight general distributions obtained collapsing $r_r$, the SW test leads to rejection of the null hypothesis of log-normality of rescaled data. This could be due to differences in upper and lower tails across distributions. To check this, and to better appreciate the effect of the rescaling process and so the strengths and weaknesses of the universality claim, Figure 1 allows comparison of the distribution of readerships (r) and rescaled readerships ($r_r$) for the sample of 2013 publications in two distinct SCs: "Cell biology" ($R_0$=62.6) and "Surgery" ($R_0$=16.2). Here, the y axis represents the probability that an article has received a number of reads at least equal to the corresponding x-axis value. While the two distributions show different patterns for original data, when rescaled they show a very similar shape (albeit only qualitatively), which "should" collapse on the black distribution represented by the lognormal that best fits the merged samples.

*Figure 1: Distribution of readership (left) and rescaled readership (right) for 2013 publications of Cell biology (yellow) and Surgery (blue)*

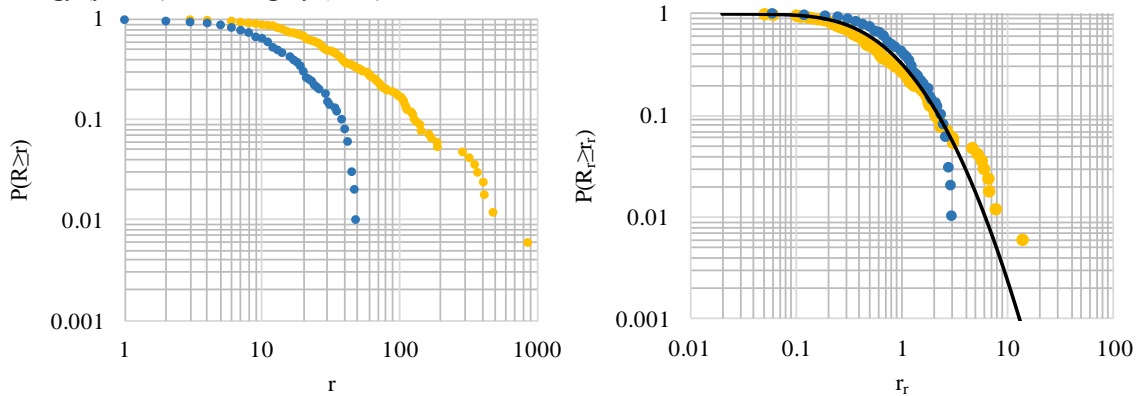

The divergence of the two curves plotted with original data is evident in the right tail of the distributions (left graph in Figure 1). The rescaling allows a good collapse of the two distributions on the lognormal curve represented by the black line in the right graph, but the upper tails continue to be well away from each other. However the example provided is that of two extreme cases: SCs with an $R_0$ ratio of 1/4. Figures 2, 3 and 4 then show the same plots for all the 30 SCs in three publication years (2010, 2013 and 2016). The left panels show the evident dispersion caused by the differing intensity of readership between fields. The rescaled distributions in the right panels tend to be less dispersed and close to the lognormal model represented by the black line, but they still show a tendency to widen in the right tail, meaning for highly read publications. It is noteworthy that the dispersion of the right tail does not seem to increase with the citation ("readership") window: Figure 4 shows in particular that for that the most recent publications, of 2016, the deviation from the universal lognormal model is due to a very limited number of publications.

---
[7] Note that all $R_0$ values are calculated "internally", i.e. on the sample data, since data on readerships for the population are unavailable.

*Figure 2: Distribution of readership (left) and rescaled readership (right) for 2010 publications (each dot series represents a subject category)*

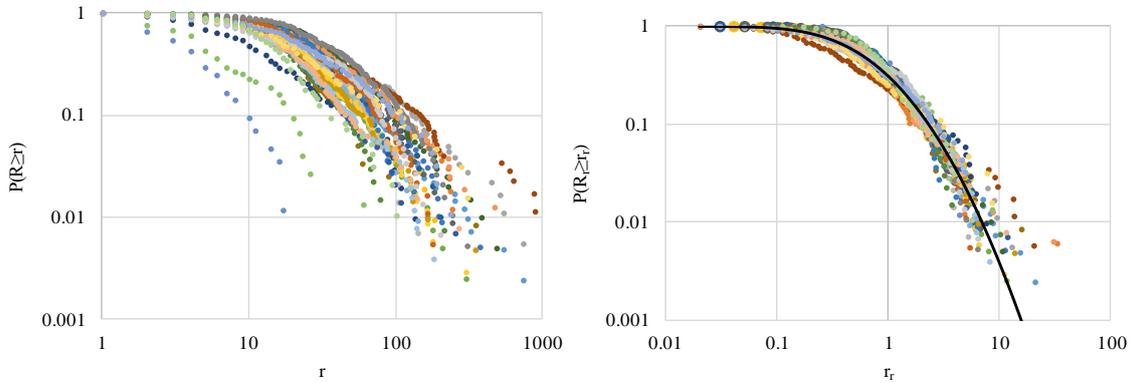

*Figure 3: Distribution of readership (left) and rescaled readership (right) for 2013 publications (each dot series represents a subject category)*

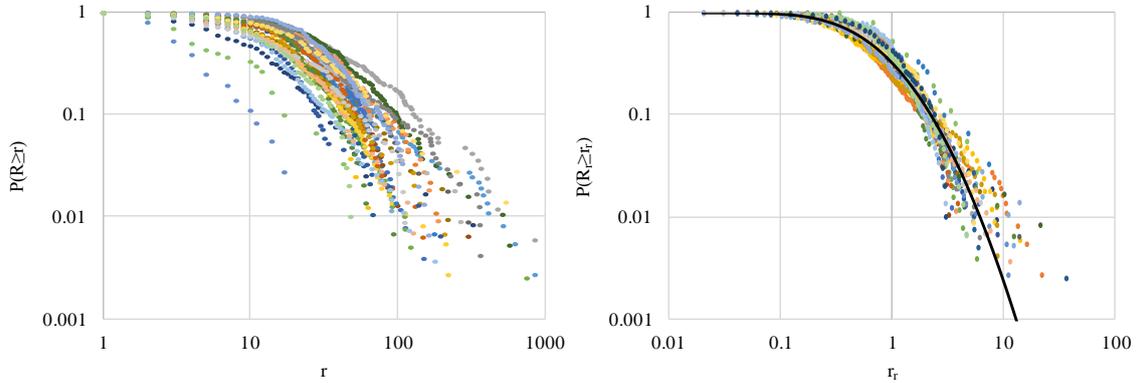

*Figure 4: Distribution of readership (left) and rescaled readership (right) for 2016 publications (each dot series represents a subject category)*

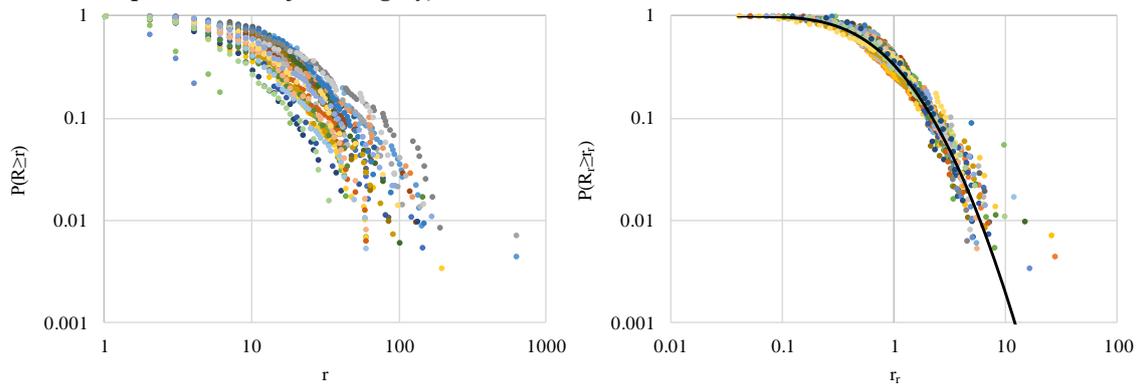

To analyze the effect of rescaling on distributions of readerships referring to different citation windows, Figure 5 shows the distribution of $r_r$, for all Biochemistry & molecular biology publications of the sample. The graph compares readership received by articles in the same subject category but published in different years and shows an effective rescaling: the dispersion around the thickening zone becomes evident only for $r_r$ greater than 5. Figure 6 then plots the dispersion of rescaled citations for the same publications, revealing, in the case of the 2016 articles (brown dots), the evident deviation from the shape. This pair of graphs confirms that: i) with a citation window of only one year, citation counts, even when rescaled, are not particularly "reliable"; ii) readerships, on the other hand, show a concurring trend for both long and short windows (consistent with prior research showing that Mendeley readers occur about a year before citations), which begins to diverge only in the final reaches, i.e. for articles read about 5/6 times the average.

*Figure 5: Distribution of rescaled readership for articles in Biochemistry & molecular biology (each dot series refers to a publication year from 2009 to 2016)*

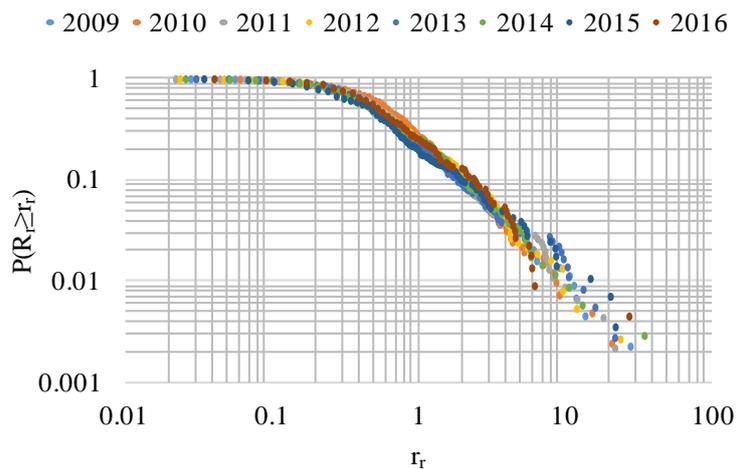

*Figure 6: Distribution of rescaled citations for articles in Biochemistry & molecular biology (each dot series refers to a publication year; brown dots represent 2016 publications)*

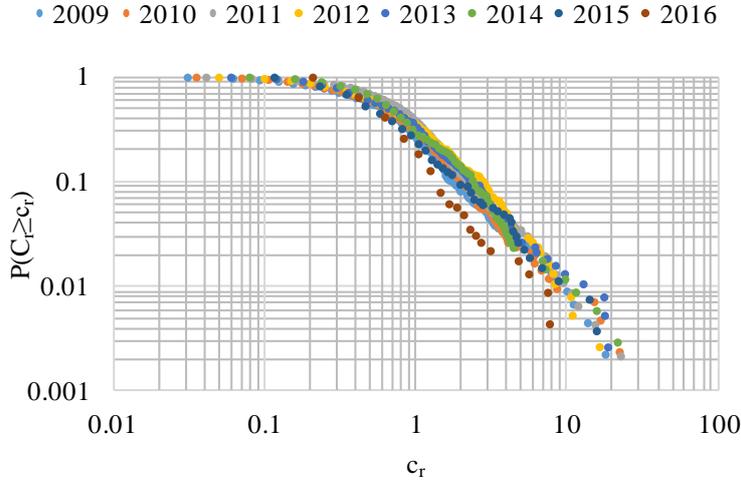

## 4.2 Effect of rescaling on top publications

As seen in the previous section, the rescaling procedure seems to confirm the same universality for Mendeley readerships as is known in the literature for citation distributions. However, there seem to be some biases in the right tails of the distributions due to the so-called extreme values. It could be useful to conduct an analysis focused on these "top read" publications. The expectation is that the rescaling would guarantee a constant share of the global top publications for all SCs and years. To this purpose, the analysis considers the articles that enter in the global top z% of ranking by number of reads.

If the distribution of the readerships were invariable across SCs, the expected percentage of top articles would be around z% in each SC, where by "around", according to Radicchi and Castellano (2011), we mean a region corresponding to the mean ± one standard deviation ($\sigma_z$), defined as:

$$\sigma_z = \sqrt{\frac{z(100-z)}{Nc} \sum_{i=1}^{Nc} \frac{1}{Ni}}$$

[3]

where:
- $N_c$ is the total number of SCs,
- $N_i$ is the total number of articles in the SC $i$.

The analysis is repeated for three values of z (top 5%, 10% and 20%), with the results shown in Figure 7, 8 and 9 for 2010 articles. Table 4 summarizes the results obtained for all publication years.

*Figure 7: Share of the 5% top-read 2010 articles in the 30 subject categories investigated, before (left) and after (right) rescaling (each bar represents a subject category; red dotted lines for the confidence interval, i.e. one standard deviation)*

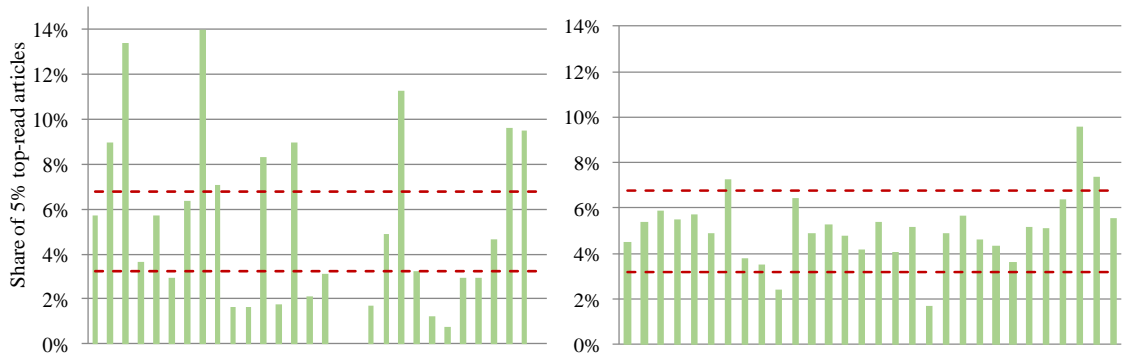

*Figure 8: Share of the 10% top-read 2010 articles, in the 30 subject categories investigated, before (left) and after (right) rescaling (each bar represents a subject category; red dotted lines for the confidence interval, i.e. one standard deviation)*

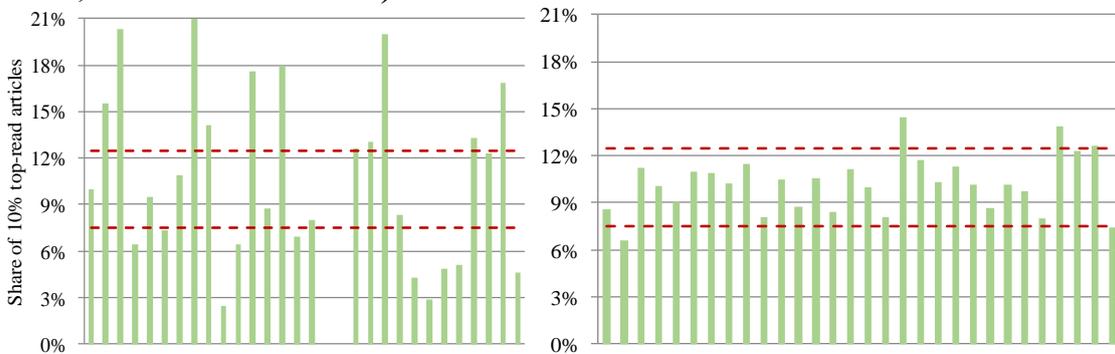

*Figure 9: Share of the 20% top-read 2010 articles, in the 30 subject categories investigated, before (left) and after (right) rescaling (each bar represents a subject category; red dotted lines for the confidence interval, i.e. one standard deviation)*

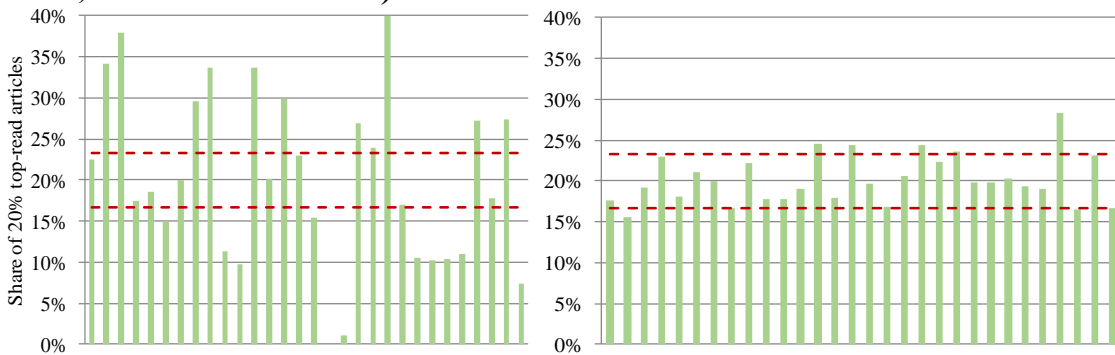

Both the graphic analysis and the numerical data of Table 4 demonstrate that rescaling produces significant effects in terms of the number of SCs in which the percentage of top publications falls within the tolerance interval. Therefore, although right tails are the most problematic concerning the universality claim, rescaling seems to allow a robust comparison even for top-read articles across different SC and publication years.

*Table 4: Number of subject categories with a share of top read articles inside the tolerance interval*

| Publication year | Top 5% | | Top 10% | | Top 20% | |
|---|---|---|---|---|---|---|
| | Original distribution | Rescaled distribution | Original distribution | Rescaled distribution | Original distribution | Rescaled distribution |
| 2009 | 11 | 25 | 9 | 25 | 11 | 20 |
| 2010 | 7 | 25 | 7 | 25 | 8 | 21 |
| 2011 | 10 | 24 | 9 | 25 | 6 | 22 |
| 2012 | 12 | 23 | 10 | 24 | 4 | 20 |
| 2013 | 5 | 23 | 6 | 26 | 5 | 20 |
| 2014 | 10 | 17 | 11 | 26 | 6 | 19 |
| 2015 | 6 | 25 | 6 | 25 | 5 | 18 |
| 2016 | 11 | 24 | 7 | 26 | 5 | 24 |

**4.3 The Characteristic Scores and Scales method**

In this section we will investigate universality for Mendeley readerships through the Characteristic Scores and Scales (CSS) method, thereby providing a different perspective in answer to the first research question presented in the introduction.

CSS involves subdividing a population into classes on the basis of the distribution of a given variable. The CSS technique was developed by Schubert, Glänzel, and Braun (1987) and has been applied in many bibliometric studies (Glänzel & Schubert, 1988; Glänzel, 2011; Ruiz-Castillo & Costas, 2014; Abramo, D'Angelo, & Soldatenkova, 2017; Bornmann & Glänzel, 2017) for classification of the given unit (publications, journals, individual scientists, research groups, institutions, etc.), in function of the value of the bibliometric indicator measured on those units.

This technique involves reiterated truncation of a frequency distribution according to mean values, also called "characteristic scores". After truncating the overall distribution at its mean value, the mean of the subpopulation above the first mean is recalculated; the subpopulation is again truncated, and so on until the procedure is stopped. Applying the CSS method to citations distribution, with up to three characteristic scores, the resulting four categories account respectively for 69-70%, 20-21%, 6-7%, 2-3% of total publications. Moreover, this breakdown into classes seems invariant with the field and the year of publication, indicating a sort of universal scheme.

Here the method will be applied to publications in the dataset to verify if the four above categories also hold for Mendeley readerships. For this purpose, we consider as characteristic scores:

$\beta_1$ = mean value of readership for the overall set of publications

$\beta_2$ = mean value for publications with readership above $\beta_1$

$\beta_3$ = mean value for publications above $\beta_2$

the procedure results in the four classes described in Table 5.

*Table 5: Classes obtained by the CSS method applied to readership of publications*

| Class | Mendeley readership thresholds | Publications |
|---|---|---|
| I | $[0, \beta_1)$ | unread or poorly read |
| II | $[\beta_1, \beta_2)$ | fairly read |
| III | $[\beta_1, \beta_2)$ | remarkably read |
| IV | $[\beta_3, \infty)$ | outstandingly read |

The application of CSS technique at overall level, i.e. to all publications in the dataset for the years 2010, 2013 and 2016, provides the results shown in Table 6[8].

*Table 6: Results from the CSS technique applied to readerships, for the overall sample of publications of years 2010, 2013, 2016*

|       | 2010      |      |       | 2013      |      |       | 2016      |      |       |
|-------|-----------|------|-------|-----------|------|-------|-----------|------|-------|
| Class | Threshold | Obs  | Share | Threshold | Obs  | Share | Threshold | Obs  | Share |
| I     | [0,28)    | 3701 | 70%   | [0,25)    | 3594 | 69%   | [0,12)    | 3413 | 66%   |
| II    | [28,69)   | 1107 | 21%   | [25,59)   | 1152 | 22%   | [12,27)   | 1218 | 24%   |
| III   | [69,134)  | 315  | 6%    | [59,120)  | 318  | 6%    | [27,49)   | 384  | 7%    |
| IV    | ≥134      | 131  | 2%    | ≥120      | 111  | 2%    | ≥49       | 157  | 3%    |

Data are in line with expectations: for readerships as well, the classic CSS distribution pattern (69-70%, 20-21%, 6-7%, 2-3%) is respected. For 2010 and 2013 the adaptation to the expected proportions is rigorous, which confirms the universal character of the distributions of this alternative metric.

For the most recent articles (2016 publications), there is a slight differentiation from the universal pattern, with a slightly lower than expected percentage of articles that are little or completely unread. This can be explained by imagining that the skewness of overall readership data, though already existent, is not yet fully formed with such a short citation ("readership") window.

The results from the analysis, repeated by stratifying the publications by SC, are shown in Table 7. In this case the reduction in the number of observations generates fluctuations in the data with respect to the expected pattern, with examples of "anomalous" minimum and maximum values. However, the analyses show a behavior that remains substantially invariant with field, confirming the "universal character" of the readerships, which was the intention of the current research.

---

[8] Results for other years are omitted for brevity, but confirm the trend emerging from the three data years considered.

*Table 7: Share of publications in each class of the CSS method, by year and subject category*

|  | 2010 | | | | 2013 | | | | 2016 | | | |
|---|---|---|---|---|---|---|---|---|---|---|---|---|
| Subject category                             Class | I | II | III | IV | I | II | III | IV | I | II | III | IV |
| Biochemistry & molecular biology | 74.0 | 19.1 | 5.3 | 1.7 | 78.5 | 16.4 | 2.7 | 2.4 | 75.9 | 17.0 | 5.6 | 1.5 |
| Biotechnology & applied microbiology | 76.0 | 18.6 | 4.8 | 0.6 | 60.8 | 28.9 | 6.2 | 4.1 | 66.7 | 23.3 | 6.7 | 3.3 |
| Cell biology | 73.3 | 19.8 | 4.8 | 2.1 | 74.3 | 19.2 | 3.6 | 3.0 | 75.1 | 17.5 | 6.8 | 0.6 |
| Chemistry, analytical | 68.8 | 22.0 | 5.5 | 3.7 | 57.5 | 24.1 | 9.2 | 9.2 | 70.0 | 22.5 | 5.0 | 2.5 |
| Chemistry, multidisciplinary | 69.5 | 21.9 | 5.2 | 3.3 | 61.2 | 23.9 | 7.8 | 7.1 | 62.4 | 22.2 | 11.5 | 3.8 |
| Chemistry, physical | 67.5 | 21.5 | 7.3 | 3.7 | 69.7 | 22.2 | 6.6 | 1.5 | 66.2 | 21.1 | 7.0 | 5.7 |
| Clinical neurology | 68.2 | 20.9 | 6.4 | 4.5 | 66.7 | 18.4 | 7.9 | 7.0 | 67.6 | 24.3 | 5.4 | 2.7 |
| Economics | 78.0 | 18.8 | 1.6 | 1.6 | 70.0 | 20.0 | 7.4 | 2.6 | 68.4 | 22.6 | 6.5 | 2.6 |
| Energy & fuels | 64.6 | 23.0 | 8.8 | 3.5 | 64.9 | 23.7 | 8.8 | 2.6 | 61.7 | 23.4 | 11.2 | 3.7 |
| Engineering, chemical | 67.7 | 24.2 | 6.5 | 1.6 | 71.8 | 20.9 | 4.5 | 2.7 | 62.4 | 24.0 | 10.4 | 3.2 |
| Engineering, electrical & electronic | 75.8 | 15.3 | 4.0 | 4.8 | 73.3 | 20.7 | 4.7 | 1.3 | 64.2 | 25.5 | 7.5 | 2.8 |
| Environmental sciences | 71.7 | 20.0 | 5.9 | 2.4 | 73.0 | 20.9 | 4.1 | 2.0 | 60.3 | 26.0 | 9.1 | 4.5 |
| Food science & technology | 64.0 | 23.7 | 7.0 | 5.3 | 57.7 | 25.8 | 10.3 | 6.2 | 65.2 | 21.7 | 8.7 | 4.3 |
| Genetics & heredity | 77.8 | 17.4 | 3.6 | 1.2 | 65.0 | 25.6 | 6.0 | 3.4 | 74.5 | 17.0 | 5.3 | 3.2 |
| Immunology | 66.0 | 20.1 | 8.3 | 5.6 | 71.3 | 22.6 | 3.5 | 2.6 | 73.5 | 20.5 | 3.4 | 2.6 |
| Materials science, multidisciplinary | 69.2 | 20.5 | 6.3 | 4.0 | 66.2 | 22.6 | 8.1 | 3.1 | 66.9 | 22.6 | 6.9 | 3.6 |
| Mathematics | 70.1 | 17.5 | 9.3 | 3.1 | 72.7 | 18.2 | 5.5 | 3.6 | 62.3 | 22.6 | 11.3 | 3.8 |
| Mathematics, applied | 67.6 | 23.0 | 5.4 | 4.1 | 63.8 | 24.6 | 5.8 | 5.8 | 64.1 | 25.6 | 9.0 | 1.3 |
| Microbiology | 64.7 | 21.8 | 10.1 | 3.4 | 64.4 | 22.1 | 8.7 | 4.7 | 65.2 | 23.0 | 8.1 | 3.7 |
| Nanoscience & nanotechnology | 67.9 | 21.2 | 7.1 | 3.8 | 61.3 | 26.3 | 8.8 | 3.8 | 67.6 | 22.1 | 4.9 | 5.4 |
| Neurosciences | 66.2 | 21.0 | 8.2 | 4.6 | 76.8 | 15.4 | 5.0 | 2.9 | 73.3 | 18.0 | 4.7 | 4.1 |
| Oncology | 70.0 | 20.3 | 6.5 | 3.2 | 70.5 | 20.8 | 6.0 | 2.7 | 69.6 | 22.2 | 4.3 | 3.9 |
| Optics | 64.8 | 25.3 | 6.2 | 3.7 | 64.0 | 29.2 | 5.0 | 1.9 | 62.5 | 25.7 | 8.8 | 2.9 |
| Pharmacology & pharmacy | 70.3 | 19.6 | 8.0 | 2.2 | 56.5 | 27.3 | 10.6 | 5.6 | 68.2 | 22.4 | 6.5 | 2.8 |
| Physics, applied | 68.0 | 23.0 | 4.9 | 4.2 | 69.2 | 20.9 | 7.2 | 2.7 | 62.2 | 25.9 | 7.8 | 4.1 |
| Physics, condensed matter | 70.1 | 22.6 | 5.1 | 2.2 | 65.8 | 21.7 | 8.6 | 3.9 | 64.0 | 23.3 | 7.3 | 5.3 |
| Physics, multidisciplinary | 65.9 | 20.8 | 8.1 | 5.2 | 66.7 | 22.5 | 6.2 | 4.7 | 69.7 | 19.3 | 6.4 | 4.6 |
| Plant sciences | 76.7 | 13.7 | 5.5 | 4.1 | 74.4 | 19.2 | 3.8 | 2.6 | 70.0 | 17.5 | 8.3 | 4.2 |
| Public, environmental & occupational Health | 70.5 | 20.0 | 5.3 | 4.2 | 65.6 | 22.7 | 6.5 | 5.2 | 70.3 | 21.8 | 5.5 | 2.4 |
| Surgery | 66.7 | 25.0 | 3.7 | 4.6 | 61.6 | 22.4 | 8.0 | 8.0 | 65.4 | 21.3 | 8.1 | 5.1 |

## 5. Conclusions

The challenge of assessing the impact of a scientific publication seems as complex as it is important. The term "impact" itself does not have a shared meaning among scholars, and the debate continues in the literature on theoretical aspects and approaches to measurement. At the same time, operative proposals and indicators are proliferating. The advent of digital technologies has made it possible to trace the spread of the many forms of new knowledge within social networks, giving rise to a new stream of analysis and of its own related literature: that of "altmetrics". The very root of this term suggests a proposal of new indicators, in counterbalance, or at least as an alternative, to "traditional" metrics. In reality, the peer review mechanism is still at the base of everything: the citations testify that the knowledge developed by a researcher (and codified in their publication) is considered important by a colleague, their peer, who in turn includes it in the bibliographies of their own works. In the same way, altmetrics attest that a given work is considered worthy of attention by another scholar, to the point of downloading it from a repository, inserting it in their own digital library, reading it, sharing it, remarking on it in social networks.

This paper has dealt with a particular type of altmetric: Mendeley readerships. Apart from the matter of a reasonable time delay (reading a publication occurs necessarily before citing it), the opening hypothesis is that this type of indicator would evidently be correlated with citations, due precisely to the unity of the underlying mechanism: peer review.

The study of the statistical properties of citations (and also of readerships) plays an important role when it comes to using them in assessing the impact of a scientific publication, whether alone or in combination.

As known, given the varying intensity of citation between fields, citation-based indicators must be standardized. The potential use of altmetrics implies the same process, which in turns calls for deeper understanding of the properties of the reference distributions.

The analyses advanced in this work, based on a random sample of WoS publications belonging to 30 subject categories, were aimed at verifying whether readership distributions share the same shape across fields and can be rescaled to a common and universal form.

Mendeley readership counts seem to follow a log-normal distribution, as in Radicchi and Castellano (2011) for citations. The application of the CSS technique confirms that readership distributions for publications in different fields share some fundamental characteristics and similarities, as stated by Albarrán et al. (2011) for citations. The rescaling process seems sufficiently effective even for top-read publications, agreeing with Abramo et al. (2012) on citations. However, the varying shape of right tails across fields presents a challenge to the universality hypothesis. Still, borrowing the words of Waltman et al. (2012), the use of the rescaled values of the indicator may be justified even when the distributions are not universal.

Continuing from this, the important news is that unlike citations, the effectiveness of rescaling readership also applies for very short citation windows: due clearly to the immediacy of the Mendeley readerships, meaning their rapid initial accumulation compared to citations. This confirms the potential for readership indicators when the assessment concerns "young" research products, as is typical of national evaluation exercises.

In conclusion, the limitations of the proposed study must be acknowledged. In fact, all the above arguments are based on an analysis carried out on a dataset limited to publications belonging to 30 subject categories, being the ones most relevant in terms of publication portfolio size. With the sole exception of Economics, social science fields are omitted, as well as all arts and humanities subject categories.